\documentclass[
    ,final            
  ]
  {aipproc}

\layoutstyle{8x11single}

\begin{document}

\title{On non-perturbative analysis of massive and bimetric gravity}


\author{Alexey Golovnev}{
  address={St. Petersburg State University; Ulyanovskaya ul. 1, 198504 St. Petersburg, Russia; agolovnev@yandex.ru}
}

\begin{abstract}

We give a brief overview of the ghost-free massive and bimetric gravity emphasising its non-perturbative aspects and ADM approach to calculating the precise number of degrees of freedom. To the standard material, we add some thoughts concerning existence and uniqueness problems of matrix square roots in non-perturbative metric formulations of massive gravity. 

\end{abstract}

\maketitle


\section{Massive gravity}

Constructing a viable theory of massive gravity \cite{FierzPauli} was a long-standing problem with the general trend of giving the negative answer \cite{BoulwareDeser}. A serious progress has been made several years ago which has led to the family of dRGT ghost-free models \cite{dRG,dRGT}. The major task was to get rid off the sixth polarisation of the graviton which is a ghost \cite{BoulwareDeser}. With modern ADM formalism, it is evident that, out of ten independent metric components, one lapse and three shifts are non-dynamical and, in absence of further dynamical constraints, all six components of the spatial metric are dynamical and independent. It exceeds the required five polarisations of a massive spin-two particle by one.

That the sixth mode is problematic was clear from the very early days of massive gravity. Indeed, the linearised Einstein-Hilbert action around Minkowski spacetime 
\begin{equation}
\label{linearEH}
S=\int d^4 x\left(-\frac14(\partial_{\alpha}h_{\mu\nu})(\partial^{\alpha}h^{\mu\nu})+
\frac12(\partial^{\alpha}h_{\mu\nu})(\partial^{\nu}h^{\mu}_{\alpha})-
\frac12(\partial_{\alpha}h^{\alpha\mu})(\partial_{\mu}h^{\beta}_{\beta})+
\frac14(\partial_{\mu}h^{\alpha}_{\alpha})(\partial^{\mu}h^{\beta}_{\beta})+{\mathcal O}(h^3)\right)
\end{equation}
in the standard perturbation variables reads
\begin{equation}
\label{pertact}
S=\int d^4 x\left(-\frac14 (\partial_{\alpha}h^{(TT)}_{ij})(\partial^{\alpha}h^{(TT)}_{ij})
+\frac12\left(\partial_j\left({\dot v}_i-s_i\right)\right)^2-6{\dot\psi}^2+2(\partial_i \psi)^2+
4\psi\bigtriangleup\left(\phi-{\dot b}+{\ddot\sigma}\right)+{\mathcal O}(h^3)\right)
\end{equation}
where as usual  $h_{00}=2\phi$, $h_{0i}
=\partial_i b+s_i$ with $\partial_i s_i=0$, and $h_{ij}=2\psi\delta_{ij}+2\partial^2_{ij}\sigma+
\partial_i v_j+\partial_j v_i+h^{(TT)}_{ij}$ with $\partial_i v_i=0$, $\partial_i h^{(TT)}_{ij}=0$
and $h^{(TT)}_{ii}=0$. 

Given the "mostly plus" signature, we see that the transverse traceless sector (two independent variables) is healthy. Two transverse vectors amount to four independent variables with only two of them, ${\dot v}_i-s_i$, gauge invariant. Finally, four scalars combine into two gauge invariant combinations, $\psi$ and $\phi-{\dot b}+{\ddot\sigma}$. For those who know we note that these gauge invariant quantities can be obtained as the ${\mathcal H}\to0$ limit of the standard gauge-invariant variables from cosmological perturbation theory in conformal time, $\Phi=\phi-{\dot b}+{\ddot\sigma}-{\mathcal H}\left(b-{\dot\sigma}\right)$ and $\Psi=\psi+{\mathcal H}\left(b-{\dot\sigma}\right)$ where  ${\mathcal H}$ is the "Hubble constant"  ${\mathcal H}\equiv\frac{\dot a}{a}$ in conformal time.

The variables $s_i$ and $\phi$ have no time derivatives in the action, and the time derivative of $b$ can be excluded by intergration by parts. Therefore, four variables are non-dynamical from the very beginning. And, due to the gauge freedom, it is easy to see that the vector and scalar sectors are fully constrained in general relativity which is extremely good because otherwise the scalar sector in the action (\ref{pertact}) contains kinetic energies of the wrong sign. Of course, the remaining helicity-two variables represent the massless graviton.

A priori, one can think of two ways of introducing a mass term into the action (\ref{linearEH}), $h_{\mu\nu}h^{\mu\nu}$ and $h^{\mu}_{\mu}h^{\nu}_{\nu}$. Unfortunately, it breaks the gauge invariance, and therefore generically makes the potentially problematic variables truly dynamical. However, we note from the kinetic term (\ref{pertact}) that $\phi$, or essentially $h_{00}$, has entered the action linearly which brings us to the Fierz-Pauli mass term
\begin{equation}
\label{FP}
V=\frac{m^2}{4}\left(h_{\mu\nu}h^{\mu\nu}-h^{\mu}_{\mu}h^{\nu}_{\nu}\right)
\end{equation}
preserving the linear dependence on $\phi$. Despite absence of gauge invariance, the field $\phi$ serves as a Lagrange multiplier entailing a constraint on dynamical sectors. One can check that five surviving degrees of freedom are healthy, provided that the sign of the mass term is correct.

Another view on the Fierz-Pauli mass term can be obtained from the St{\" u}ckelberg trick, see a nice detailed discussion in the review paper \cite{Kurt}. At linear order, one can mimic the gauge transformation by
$$h_{\mu\nu}\to h_{\mu\nu}+\partial_{\mu}\xi_{\nu}+\partial_{\nu}\xi_{\mu}$$
with an auxiliary 4-vector $\xi_{\mu}$. Substituting it into the mass term (\ref{FP}), we see that the Fierz-Pauli combination is special in that, up to surface terms, it gives the Maxwellian kinetic function for $\xi_{\mu}$
$$\left(\partial_{\mu}\xi_{\nu}-\partial_{\nu}\xi_{\mu}\right)\left(\partial^{\mu}\xi^{\nu}-\partial^{\nu}\xi^{\mu}\right)$$
which is healthy. Otherwise the subsequent St{\" u}ckelberg trick $$\xi_{\mu}\to\xi_{\mu}+\partial_{\mu}\varphi$$ would have produced a higher-derivative action for $\varphi$. This is how the helicity zero mode appears to have two degrees of freedom with different signs of kinetic energy.

The problem with this discussion is that it is entirely in terms of the linearised theory. Going beyond that, we have to take into account the quadratic in $\xi_{\mu}$ part of the St{\" u}ckelberg transformation which destroys the Maxwellian form of the vector kinetic function. Effective field theory breaks down at the $\Lambda_5\equiv \left(m^4 M_{Pl}\right)^{1/5}$ scale, with the leading terms in the quartic self-interaction of $\varphi$. The basic idea of dRGT \cite{dRG} was to amend the model first by a cubic addition to the Fierz-Pauli mass term, then to correct the new leading scalar self-interaction by the next order contribution to the potential, and so on. It allows to push the cutoff scale up to $\Lambda_3\equiv \left(m^2 M_{Pl}\right)^{1/3}$, and there were good reasons to suspect that, at least classically, the model gets restricted to precisely five independent degrees of freedom.

\section{Non-perturbative definition and analysis}

It has been noticed \cite{dRGT,HR1} that the series expansion of the mass term was nothing but Taylor expansion of a square root, so that
\begin{equation}
\label{dRGT}
V=2m^2\left( {\rm Tr}\sqrt{g^{\mu\alpha}\eta_{\alpha\nu}}-3\right)
\end{equation}
where $\eta_{\mu\nu}$ is Minkowski metric, and the $-3$ term serves only to keep the Minkowski space a solution.
The non-perturbative calculation of the number of degrees of freedom required the ADM analysis to be used. General relativity with an additional potential $V$ has the Hamiltonian
\begin{equation}
\label{Hamiltonian}
H=-\int d^3 x\sqrt{\gamma}\left(N\left(\mathop{R}\limits^{({\mathit 3})}+\frac{1}{\gamma}\left(\frac12 \left(\pi^j_j\right)^2-\pi_{ik}\pi^{ik}\right)
-V\right)+2N^i\mathop{{\bigtriangledown}^k}\limits^{({\mathit 3})}\pi_{ik} \right)
\end{equation}
where the ADM variables are given by
\begin{equation}
\label{ADM}
ds^2\equiv g_{\mu\nu}dx^{\mu}dx{\nu}=-(N^2-N_k N^k)dt^2+2N_i dx^i dt+\gamma_{ij}dx^i dx^j.
\end{equation}
$N$ and $N_i$ are the lapse and shift functions respectively, and $N_i\equiv\gamma_{ik}N^k$. Generically the potential term non-linearly depends on all the lapse and shifts, and therefore the variation with respect to those variables directly determines their values in terms of the spatial metric components instead of producing any constraints on $\gamma_{ij}$.

The non-perturbative analysis has first been done \cite{HR2} by directly square-rooting the matrix
\begin{eqnarray}
\label{basic}
g^{\mu\alpha}\eta_{\alpha\nu}=
\left( \begin{array}{cc}
\frac{1}{N^2} & \frac{N^i}{N^2} \\
-\frac{N^j}{N^2} & \gamma^{ij}-\frac{N^i N^j}{N^2} 
\end{array} \right).
\end{eqnarray}
By a complicated direct calculation, it has been shown that, after a suitable redefinition of shifts of the form
$$N^i=\left(\delta^i_j+ND^i_j(\gamma, n)\right)n^j,$$
the square root can be expressed as
\begin{equation}
\label{decomposition}
\sqrt{g^{-1}f}=
\frac{1}{N\sqrt{1-n^kn^k}}
\left( \begin{array}{cc}
1 & n^i \\
-n^j & -n^i n^j 
\end{array} \right)+
\left( \begin{array}{cc}
0 & 0 \\
0 & X^{ij}(\gamma,n) 
\end{array} \right).
\end{equation}
Being multiplied by the measure factor $\sqrt{g}=N\sqrt{\gamma}$ in the Hamiltonian (\ref{Hamiltonian}), it becomes a linear function of the lapse, and so a spatial sector constraint emerges. Rather laborious calculations \cite{HRfull} show that this constraint produces a secondary constraint, and this pair of second class constraints allows to consistently exclude one degree of freedom, and finally an equation for the lapse can be obtained. It gives a fully self-contained dynamical system with five degrees of freedom.

In our paper \cite{my} an alternative approach was proposed. We introduce a matrix of auxiliary fields $\Phi^{\mu}_{\nu}$ and substitute the potential (\ref{dRGT}), for simplicity without the $-3$ term, by
\begin{equation}
\label{our}
V=\frac{m^2}{N}\left( \Phi^{\mu}_{\mu}+ \left(\Phi^{-1}\right)^{\mu}_{\nu}N^2 g^{\nu\alpha}\eta_{\alpha\mu}\right).
\end{equation}
Obviously, we have to assume $\Phi_i^k=\Phi^i_k$ and $\Phi^0_i=-\Phi^i_0$. After that the $\Phi$ fields can be integrated out giving $\Phi^2=N^2g^{-1}\eta$, and the initial action is restored. Associated with absence of velocities for $N$, $N^i$ and $\Phi^{\mu}_{\nu}$, we have constraints on the variables of the model. 

If there were six degrees of freedom, one would be able to directly determine all these non-dynamical variables in terms of $\gamma_{ij}$ with only these constraints. However, there is a direction in space of these variables along which there is no restriction at this level of Hamiltonian analysis \cite{my}. It signals a spatial sector constraint. In this approach we do not need to find the square root explicitly, and the calculations are straightforward even if cumbersome. Auxiliary fields appeared to be convenient for performing the non-perturbative analysis in St{\" u}ckelberg variables \cite{HRStueck}.

\section{Generalisations}

The most immediate generalisation is to consider a curved auxiliary metric
\begin{eqnarray}
\label{ADMinv2}
f_{\mu\nu} = 
\left( \begin{array}{cc}
-\left( M^2-M_k M^k\right) & M_i \\
M_j & s_{ij}
\end{array} \right)
\end{eqnarray}
 instead of Minkowski background. It suffices to substitute $\eta_{\mu\nu}$ with $f_{\mu\nu}$ in the potential term and to take the square root of the new matrix
\begin{eqnarray}
\label{basic2}
g^{\mu\alpha}f_{\alpha\nu}=
\left( \begin{array}{cc}
\frac{M^2-M_k \left(M^k-N^k\right)}{N^2} & \frac{s_{ij}N^j-M_j}{N^2} \\
-\frac{N^j\left(M^2-M_k \left(M^k-N^k\right)\right)}{N^2}+\gamma^{ij}M_{j} & 
s_{ik}\gamma^{kj}-\frac{s_{ik}N^k N^j-N_i M^j}{N^2} 
\end{array} \right).
\end{eqnarray}
The same analysis as above goes through with some mild complications \cite{HRcurved}.

One more generalisation is to note that, in four dimensions, there are three families of ghost-free massive terms \cite{HR1}:
$$V_1={\rm Tr}\sqrt{g^{-1}f},$$
$$V_2=\left({\rm Tr} \sqrt{g^{-1}f}\right)^2-{\rm Tr} \left(\sqrt{g^{-1}f}\right)^2,$$
$$V_3=\left({\rm Tr} \sqrt{g^{-1}f}\right)^3-3\left({\rm Tr} \sqrt{g^{-1}f}\right)\left({\rm Tr} \left(\sqrt{g^{-1}f}\right)^2\right)
+2{\rm Tr} \left(\sqrt{g^{-1}f}\right)^3,$$
all being symmetric polynomials of eigenvalues of $\left(\sqrt{g^{-1}f}\right)^{\mu}_{\nu}$. Moreover, one can consider massive gravity in arbitrary spacetime dimensions and use the full series $V_1, \ldots, V_{D-1}$ of elementary symmetric polynomials as ghost-free potentials. The proof of ghost-freedom with explicit square root (\ref{decomposition}) is not difficult.

We would only mention that there are further generalisations available. One can write down an independent Einstein-Hilbert term for the metric $f_{\mu\nu}$, and the Boulware-Deser ghost is still absent because the elementary symmetric polynomials of $\left(\sqrt{g^{-1}f}\right)^{\mu}_{\nu}$-eigenvalues multiplied by $\sqrt{-g}$ can obviously be expressed by means of elementary symmetric polynomials of $\left(\sqrt{f^{-1}g}\right)^{\mu}_{\nu}$-eigenvalues multiplied by $\sqrt{-f}$. Note that the last elementary symmetric polynomial $V_D$ introduces a cosmological term for the $f_{\mu\nu}$-metric. And in the vielbein formalism this construction can naturally be generalised to multimetric models \cite{KR}.

Finally, one can make the graviton mass a function of a scalar field \cite{varmass}, $V=2m^2(\sigma)\cdot\left(V_1+\sum\limits_{i=2}^{D-1} \alpha_i V_i\right)$. Alternatively, a scalar field can be incorporated in a more involved way known as quasi-dilaton massive gravity \cite{qdil}. And $f(R)$-type generalisations of massive gravity are also known \cite{fRmass}.

\section{The issue of square roots}

There is an interesting problem in non-perturbative metric formulation of massive gravity \cite{Cedric1,Cedric2,myrev}. A real square root of $\left(g^{-1}f\right)^{\mu}_{\nu}$ is not always available. In perturbation theory one can never encounter this problem. Non-perturbatively, the necessary and sufficient condition for existence of a real square root is that either $\left(g^{-1}f\right)^{\mu}_{\nu}$ does not have real negative eigenvalues, or if there is one, then it must go with an even number of identical Jordan blocks. 

What is probably more important, there is a non-uniqueness problem. Even for a unit marix one can find infinitely many real square roots like the following one:
\begin{equation}
\left( \begin{array}{cc}
3/5 & -4/5 \\
-4/5 & -3/5
\end{array} \right)\cdot
\left( \begin{array}{cc}
3/5 & -4/5 \\
-4/5 & -3/5
\end{array} \right)=
\left( \begin{array}{cc}
1 & 0 \\
0 & 1 
\end{array} \right)
\end{equation}
which works simply because $3^2+4^2=5^2$. What to make out of this infinite degeneracy?

In principle, one can use the standard theory of matrix functions. Those can be approached via interpolation polynomials, or using the Jordan normal form to define any desired function. For diagonalisable matrices the procedure is obvious, and for a $k\times k$ Jordan block we take
\begin{equation}
f\left(
\left( \begin{array}{cccc}
 \lambda  & 1  &  & \\
 & \lambda & \ddots  &   \\
 &  & \ddots  &  1 \\
 &  &   &  \lambda 
\end{array} \right)\right)\equiv
\left( \begin{array}{cccc}
 f(\lambda)  & f^{\prime}(\lambda)  & \ldots  & \frac{f^{(k-1)}(\lambda)}{(k-1)!} \\
 & f(\lambda) & \ddots  & \vdots  \\
 &  & \ddots  &  f^{\prime}(\lambda) \\
 &  &   &  f(\lambda)
\end{array} \right)
\end{equation}
The problem with this definition is that sometimes it yields complex results while a real square root also exists. The simplest example is the $\left( \begin{array}{cc}
-1 & 0 \\
0 & -1 
\end{array} \right)$ matrix which admits a real square root $\left( \begin{array}{cc}
0 & -1 \\
1 & 0 
\end{array} \right)$. In other words, the problem comes with real negative eigenvalues, in this case with two identical $1\times 1$ Jordan blocks. Note though that even with another square root, $\left( \begin{array}{cc}
i & 0 \\
0 & i 
\end{array} \right)$, the potential $V_2$ would still be real-valued\footnote{Fawad Hassan, private communication}.

Of course, this problem arises only in non-perturbative regimes when the two metrics are in a sense very far from each other. A real negative eigenvalue of $\left(g^{-1}f\right)^{\mu}_{\nu}$ obviously means that there exists a positive linear combination of $g_{\mu\nu}$ and $f_{\mu\nu}$ which is degenerate \cite{myrev}. At the same time, for complex eigenvalues no such problem presents itself. Indeed, for a real-valued matrix, complex eigenvalues go in complex-conjugate pairs with complex-conjugate eigenvectors, and after transforming back to the initial basis, they will result in a real square root, provided that the branches of $\sqrt{\lambda}$ are chosen consistently for the two complex-conjugate eigenvalues.

In principle, one can also try using Frobenius normal forms over real numbers for defining the square roots. However, the proof of ghost-freedom goes through even with auxiliary fields $\Phi$ which probably means that the structure of the model is insensitive to the way of taking the square root. It is still an open problem to understand the algebraic peculiarities behind the massive gravity actions.

\section{Conclusions}

General relativity is a very rich theory. And, upon any intrusions, its scalar sector can present many unexpected surprises. For example, somehow modifying the Einstein-Hilbert variational principle in its conformal part, one can have an arbitrary cosmological constant without interactions with the trace part of the stress tensor as in unimodular gravity \cite{unimgr}, or be able to mimic the Dark Matter behaviour with modified gravitation \cite{MDM,myMDM,MDM2}. Analysing the conformal sector in a gravitational model with several new degrees of freedom, it is sometimes possible to prove that it is healthy \cite{Tomi} despite the ghosts in other modes \cite{weall}. And vice versa, a single scalar ghost can spoil the whole game in otherwise very promising model as it was with the Boulware-Deser mode of massive graviton. And among all these blessings, deceptions, and dangers, it is the dangerous and even fatal part which appears to be the most exciting one. Fighting with the sixth polarisation has produced a very interesting model, the dRGT gravity. We have reviewed its formal aspects totally neglecting phenomenological questions for which we refer the reader to a very recent review by Claudia de Rham \cite{Claudia}, one of the founders.

\begin{theacknowledgments}
  The author was partially supported by Saint Petersburg State University research grant No. 11.38.660.2013 and by Russian Foundation for Basic Research grant No. 12-02-31214.
\end{theacknowledgments}

\bibliographystyle{aipproc}   

\end{document}